\documentclass[journal,a4paper]{IEEEtran}

\usepackage[utf8]{inputenc}

\usepackage{cite}
\usepackage{graphicx}
\usepackage[dvipsnames]{xcolor}
\graphicspath{{./figures/}}

\usepackage[export]{adjustbox}

\newlength{\myfillwidth}
\newlength{\myfillheight}
\newcommand{\includegraphicsfilled}[3]{
\setlength{\myfillwidth}{#1}
\setlength{\myfillheight}{#2}
\adjustbox{%
min size={\myfillwidth}{\myfillheight},%
Clip*={.5\width-0.5\myfillwidth} {\totalheight-\myfillheight} {.5\width+0.5\myfillwidth} {\totalheight}}{%
\includegraphics[max size={\myfillwidth}{\myfillheight}]{#3}}
}

\newcommand{\figref}[1]{\figurename~\ref{#1}}
\newcommand{\secref}[1]{Section~\ref{#1}}
\usepackage{xspace}
\newcommand{\eg}[0]{e.\,g.\xspace}

\usepackage{hyperref}

\begin{document}
\bstctlcite{IEEEexample:BSTcontrol}

\title{CAI4CAI: The Rise of Contextual Artificial Intelligence in Computer Assisted Interventions}

\author{Tom~Vercauteren,
        Mathias~Unberath,
        Nicolas~Padoy,
        and~Nassir~Navab%
\thanks{T. Vercauteren is with the School of Biomedical Engineering {\&}  Imaging Sciences, King's College London.}%
\thanks{M. Unberath is with the Department of Computer Science, Johns Hopkins University.}%
\thanks{N. Padoy is with the ICube institute, University of Strasbourg, CNRS, IHU Strasbourg, France}%
\thanks{N. Navab is with the Fakult\"at f\"ur Informatik, Technische Universit\"at M\"unchen.}%
}

\markboth{Manuscript published in the Proceedings of the IEEE}%
{Contextual AI for Computer Assisted Interventions}

\maketitle

\begin{abstract}
Data-driven computational approaches have evolved to enable extraction of information from medical images with a reliability, accuracy and speed which is already transforming their interpretation and exploitation in clinical practice. While similar benefits are longed for in the field of interventional imaging, this ambition is challenged by a much higher heterogeneity. Clinical workflows within interventional suites and operating theatres are extremely complex and typically rely on poorly integrated intra-operative devices, sensors, and support infrastructures.
Taking stock of
some of the most exciting developments in machine learning and artificial intelligence for computer assisted interventions, we highlight the crucial need to take context and human factors into account in order to address these challenges.
Contextual artificial intelligence for computer assisted intervention, or CAI4CAI, arises as an emerging opportunity feeding into the broader field of surgical data science. 
Central challenges being addressed in CAI4CAI include how to integrate the ensemble of prior knowledge and instantaneous sensory information from experts, sensors and actuators; how to create and communicate a faithful and actionable shared representation of the surgery among a mixed human-AI actor team; how to design interventional systems and associated cognitive shared control schemes for online uncertainty-aware collaborative decision making ultimately producing more precise and reliable interventions.
\end{abstract}


\section{Introduction}%
\label{sec:intro}
\IEEEPARstart{C}{ontemporary} progresses in machine learning and artificial intelligence have permitted the development of tools that can assist clinicians in exploiting and quantifying clinical data including images, textual reports and genetic information. State-of-the-art algorithms are becoming mature enough to provide automated analysis when applied to well-controlled clinical studies and trials~\cite{Simpson:arXiv:2019,Gibson:CMPB:2018}, but adapting these tools for patient-specific management remains an active research area, with the bulk of the research community having focused on fully automated machine learning tools. These considerations become especially critical in the highly heterogeneous context of surgery and interventional procedures which require patient- and team-specific decision support tools able to draw information from non-standardised interventional devices integrated in diverse interventional suites.
Compared to computational tasks in radiology, the domain of computer-assisted intervention further creates unique methodological challenges, such as imposing stringent time constraints in the interventional suite, requiring knowledge of procedural data, and needing methods that deal with dynamic environments.

In this paper, keeping a focus on imaging data, we review existing work, and share insights on future developments, of machine learning strategies that decipher, support, augment and integrate in various surgical and interventional workflows while providing the flexibility required by clinical management.
Flexibility is for example mandated to be able to deal with missing input sources, react to real-time user feedback, adapt to the patient risk aversion and preferences, handle uncertain or contradictory information, learn from potentially small and heterogeneous data, etc. All of which are common in computer assisted interventions.
Imaging sources of particular interest for surgery and intervention include a wide range of well-known interventional modalities such as surgical microscopy, video endoscopy, X-ray fluoroscopy and ultrasound; more emerging biophotonics imaging modalities such as hyperspectral imaging, endomicroscopy and photoacoustic imaging; but also span classical radiology modalities such as MRI and CT which remain the main sources of imaging data for pre-operative intervention planning and post-operative assessment.
We argue that the stringent need to consider context when analysing surgical and interventional data coupled with the heterogeneity of information sources and domain knowledge in computer assisted intervention applications calls for the development of novel domain-specific contextual artificial intelligence solutions, a domain we coin as CAI4CAI.
Feeding into the broader field of Surgical Data Science~\cite{Maier-Hein:NATBME:2017,Vedula:ISS:2017,Maier-Hein:arXiv:2018}, CAI4CAI will focus on the underpinning machine learning methodology exploiting contextual information and human interaction to enable the required responsiveness to deliver clinical impact in surgery and interventional sciences.

To support our claim, we highlight some of the transformative machine learning research results and methodologies currently being developed across the spectrum of tasks in computer assisted interventions.
The impact of machine learning in intervention planning is discussed in \secref{sec:planning};
intra-operative data fusion in \secref{sec:fusion};
intelligent intra-operative imaging in \secref{sec:modality};
surgical and endoscopic vision in \secref{sec:surgvision}; and
clinical workflow monitoring and support in \secref{sec:datasci}.
In these sections, we will highlight how flexible deep-learning based tools are becoming critical for the design of effective and efficient intervention planning solutions.
During surgery, navigation solutions are often used to map preoperative information in the context of the intervention. However, navigation does not account for intra-operative changes. Learning how to co-register images is now leading to intra-operative registration solutions that are able to cope with the highly challenging task of aligning pre-operative to intra-operative images coming from different imaging modalities.
Concurrently, AI methodology is advancing to go beyond traditional navigation-based data fusion and image overlay to exploit information coming from complex or synergistic data sources. This is giving rise to what we refer to as intelligent intra-operative imaging.
Data-driven modelling strategies coming from the computer vision community are acting as instrumental starting points to achieve semantic information extraction from interventional data sources including endoscopic videos, with applications ranging from automated polyp detection to surgical activity recognition.
To deliver improved clinical outcomes through AI, all these building blocks are increasingly being integrated at the level of the complete surgical workflow with applications spanning the full breath of surgical data science. In this area, starting from data-driven mapping of clinical workflow and skills assessment, AI is now helping make contextual decision support tools and conditionally autonomous intervention a reality.
Finally, closing thoughts are provided and further budding applications of CAI4CAI are discussed in \secref{sec:outlook}.

\section{Intervention Planning}
\label{sec:planning}

\subsection{Clinical Adoption of Intervention Planning Tools}
\label{sec:planningadoption}
Once a decision is made for a patient to undergo an interventional procedure, for any non-trivial operation, patient-specific planning of the intervention is required.
The steps involved usually necessitate acquisition of reference pre-operative imaging data, semantic segmentation of anatomical structures in these images, determination of the surgical approach and elaboration of an intra-operative plan leading to optimal outcomes for the patient.
Such a plan might encompass establishing a surgical path and target, designing or selecting a patient-specific implant or assistive adjunct tool such as a drill or saw guide \cite{Peters:Book:2008}.
In the vast majority of cases, such intervention planning is performed by a team of healthcare professionals, each with their own expertise, known as the multidisciplinary team (MDT).
Relatively little computer assistance is currently available for interventional planning in the clinic.
Notable exceptions can be found in the field of neurosurgery, oral and maxillofacial surgery and orthopedic surgery.
What these specialties share is a relatively static surgical scene thanks to the proximity of rigid bone structures.
Computed tomography (CT) provides a rich source of 3D imaging information in this context. Indeed, thanks to the quantitative nature of CT images and the good contrast of bone, automated segmentation of bone has proven to be clinically reliable. Since the seminal work of the Retrospective Registration Evaluation Project (RREP) \cite{West:JCAT:1997}, it is also clear that preoperative rigid registration of different imaging modalities such as MR and CT provides a robust means of fusing soft tissue contrast information with accurate bone delineation for neurosurgical planning.
Such technical advances have supported the adoption of stereotactic surgery as a means of accurately targeting  and guiding instrument towards deep seated brain structures for procedures such as brain biopsies for tumour grading and electrode implantation for the treatment of movement disorder or the localisation of epileptic seizure onset zones.
While computer assisted surgical planning and subsequent surgical navigation have become standard of care in neurosurgery and a few other disciplines, even in these fields, there is major scope to make the workflow more efficient through the development of further machine learning enabled computer assistance.

\begin{figure}[tb]
\centering
\includegraphics[width=\linewidth]{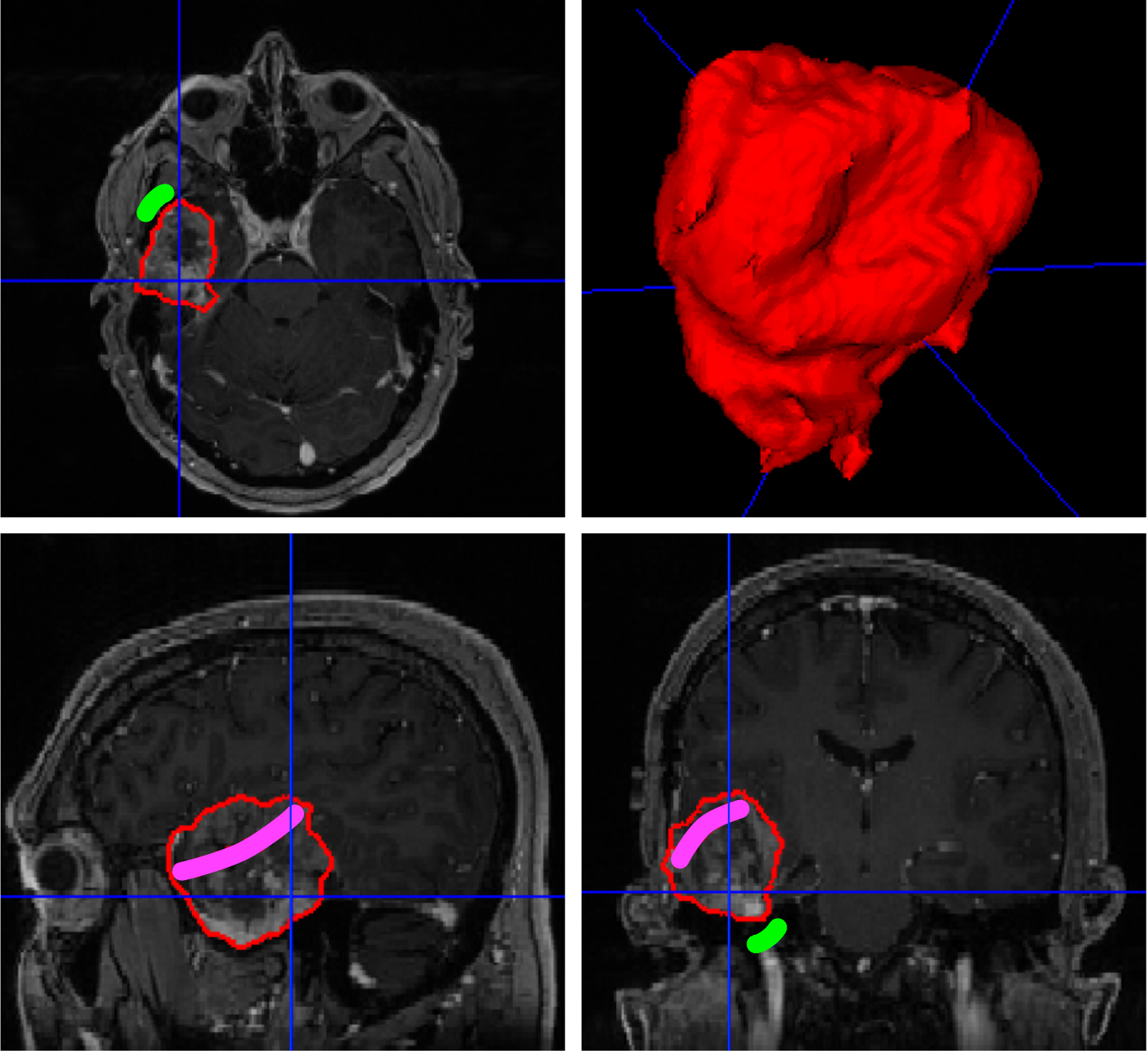}
\caption{Interactive algorithms are required to deliver context-aware artificial intelligence. In this example, using the algorithm presented in \cite{Wang:PAMI:2019}, brain tumour segmentation is initially performed automatically using a pre-trained algorithm. As part of the surgical planning, the user may want to refine the segmentation by providing scribbles to denote areas that should be excluded (green) or included (pink) irrespective of the initial segmentation. The algorithm then adapts its output to respect the user input.}
\label{fig:InteractiveSeg}
\end{figure}

\subsection{Machine Learning in Interventional Planning}
\label{sec:mlinplanning}
Commercial surgical planning products are still limited in the automation they support, with many of the most advanced ones essentially relying on classical image analysis methods such as atlas-based segmentation \cite{Iglesias:MedIA:2015} to delineate soft-tissue structures of interests for patient showing no gross pathological brain changes. Clinicians are often left with manual or generic interactive methods to delineate other structures of interest and define their surgical plan.
When interventional planning only relies on the clinician getting a volumetric representation of the patient anatomy from pre-operative data, advanced visualisation techniques such as cinematic rendering~\cite{Comaniciu:MedIA:2016} can be considered as alternatives to explicit segmentation of structures. These may produce results that are less sensitive to noise and data variability but do not enable more quantitative planning.
Developments of deep machine learning segmentation algorithms dedicated to medical imaging \cite{Li:IPMI:2017,Litjens:MedIA:2017} is rapidly changing to level of accuracy at which automated segmentation of structures of interest can be done in population of patients even in the presence of gross pathological changes \cite{Bakas:arXiv:2018}.
Yet many challenges remain for these tools to become of practical use for intervention planning purposes.
Poor generalisation when faced with slight domain changes is a recognised problem in the entire medical imaging community including on the diagnostic side.
Expanding the size of the datasets on which deep learning algorithms are trained would certainly mitigate generalisation issues by providing a much larger variety of training cases.
Collaborative efforts within the community are notably focusing on providing open-access large annotated datasets for machine learning training purposes in some specific use cases \cite{Simpson:arXiv:2019}.
Yet, collecting task-specific large annotated databases for medical imaging purposes faces its own challenges given the time and expertise required to provide detailed annotations as well as the legal, privacy and storage questions pertaining to sharing large patient datasets across multiple sites. Federated learning for multi-institutional collaboration in medical imaging \cite{Sheller:BrainLes:2018,Roy:arXiv:2019} provides a potential technical solution to this problem. Implementing such solutions at scale will require concerted efforts reaching far beyond the methodological research community. Furthermore, changes such as device upgrades or challenges posed by new clinical indications will not be captured by increasing the pool of retrospective training data. Active research to address such inevitable but unpredictable domain gaps is rooted in domain adaptation techniques \cite{Kamnitsas:IPMI:2017}.
These advances are necessary for automated machine learning tools to make an impact in the clinical setting. Prospective randomised clinical trials (RCT) are widely seen as the only source of trustworthy clinical evidence, yet studies  implementing RCTs with systems relying on deep learning tools for medical imaging currently remain noteworthy exceptions~\cite{Wang:GUT:2019}.

\subsection{Importance of Flexible Contextual Machine Learning}
\label{sec:contextual planning}
What distinguishes segmentation in surgical planning from segmentation in diagnostic imaging is nonetheless that the objective is not necessarily always that of reaching the best performance in getting the structures delineated with sub-voxel accuracy. Surgical planning needs to respect patient-specific needs and preferences of the surgeon. This requires putting the clinical team at the centre and promoting flexible tools that integrate into the surgical workflow.
Interactive deep learning methodologies are emerging to combine rich prior knowledge embedded in retrospective data from previous patients with as-sparse-as-possible annotations provided by clinicians \cite{Wang:TMI:2018,Wang:PAMI:2019}. As illustrated in \figref{fig:InteractiveSeg}, deep interactive segmentation allows the clinical expert to refine the results from an initial automated step and most importantly to adapt the inferred results on the fly based on contextual information.
Furthermore, given the heterogeneity and evolving nature of surgical practice, additional flexibility is required to handle potentially missing input modalities. Recent work in deep machine learning are focusing on dealing with such dynamic hetero-modal context while exploiting heterogeneous sources of data for the training process \cite{Dorent:MIDL:2019,Dorent:MICCAI:2019}.
Bringing flexible machine learning tools to maturity will certainly play an important role in supporting the clinical adaption of AI in surgery.

As highlighted above, segmentation of structures from pre-operative images is often the foundation of computer assisted surgical planning and this currently remains the state-of-the-art in many commercial solutions. Such static segmentation when combined with intra-operative registration already provides useful surgical navigation information for relatively static surgical scenes as is the case in neurosurgery.
Nevertheless, computer assistance for intervention planning has the potential to provide impact much beyond the ability to automate the creation of 3D anatomical models and overlay of functional data.
Patient-specific simulation of given surgical plans has for example been introduced in orthopedic surgery with a long history in acetabular fracture surgery \cite{Digioia:OpTechOrtho:2000}.
State-of-the-art orthopedic surgery planning systems allow to design patient-specific implants and patient-specific surgical guides by enabling the simulation of the effect of different implants and implantation strategy on key outcome-related parameters such as the range of motion of an articulation or the limb length~\cite{Boudissa:ExpRevMedDev:2018}.
Yet, these tools often ignore the effect of soft-tissue in the simulation process and still require very labour intensive work for the surgical team to design patient-specific plans.
Experts systems capable of automatically optimising the surgical plan for a given orthopedic surgery are now being developed \cite{Kulyk:MSKI:2018} and promise to make surgical planning more efficient \cite{Carrillo:MICCAI:2017}.
In the context of deep brain insertion of instruments, machine learning approaches capable of automatically planning trajectories of multiple instruments, to maximise the efficacy of the surgery while minimising intra-operative risks and avoiding collisions between instruments have demonstrated a significant reduction in planning time for the implantation of stereo-electroencephalography electrodes for epilepsy treatment \cite{Sparks:IJCARS:2017} and for laser  interstitial  thermal  therapy \cite{Vakharia:Epilepsia:2018}.
Contextual and flexible machine learning for surgical planning promises to push the boundaries of interventional planning by exploiting data-driven approaches and real-time user feedback to efficiently plan for complex situations.
An instrument bending model was for example trained in \cite{Granados:MICCAI:2018} to predict the deviation between an original surgical plan assuming rigid electrodes and the actual electrode paths as measured on a post-operative CT. Provided reliable uncertainty estimates on the prediction can be achieved, embedding such deflection models in the trajectory planning is expected to improve the safety and accuracy of stereo-electroencephalography electrode implantation planning.

Effectively, planning is moving away from extraction of information captured in existing data and representative of a given (pre-operative) time point. Context-aware learning methods are now being developed to also predict therapy-related changes and better inform interventional planning.
By exploiting computationally complex noninvasive cardiac electrophysiology modelling coupled with transfer learning approaches,
\cite{Giffard-Roisin:TBME:2018} notably achieved online personalized predictions of electrophysiology cardiac resynchronization therapy responses, thereby paving the way for better patient selection and patient-specific therapy optimisation.
In non-quasi-static environments, surgical planning is currently further limited by our capabilities to predict intra-operative anatomical changes. In abdominal surgery for example, segmentation of structures from preoperative images may inform the clinician about the relative spatial organisation of lesions and vascular structures. However, at the onset of a minimally invasive procedure, gas insufflation is typically performed to create the surgical workspace. This has a serious impact on the geometry of the anatomy and challenges any attempt of intra-operative use of a 3D model of the anatomy generated from pre-insufflation images.
Current approaches typically rely on focusing on smaller regions where rigidity assumptions between pre- and intra-operative data may still hold \cite{Ramalhinho:IJCARS:2018} thereby limiting the scope of the surgical planning.
Data-driven prediction of anatomical changes relating to gas insufflation in laparoscopic surgery was proposed in \cite{Johnsen:MICCAI:2015}.
Still in the context of liver surgery, a system able to take into account non-imaging patient data and factual knowledge gathered from quotable sources such as clinical guidelines was proposed to support individualised treatment planning \cite{Marz:IJCARS:2015}. While relying on handcrafted features and exploiting models with limited expressiveness, this study paved the way for more holistic interventional planning.
It is expected that context-aware interventional planning will be informed by refined prediction models to suggest therapeutic plans cognizant of clinical experience as well as potential intra-operative changes and associated risks but also flexible enough to take into account any further input from the interventional team interacting with a responsive planning system.

\section{Intra-operative Data fusion}
\label{sec:fusion}

\subsection{Navigation and Image Registration Challenges}
No matter how refined and capable interventional planning becomes, its full value
for procedural guidance and intra-operative decision making support remains contingent on appropriate geometric alignment with intra-operatively acquired data.
This alignment is achieved using registration methods that either rely on dedicated external hardware, such as optical or electromagnetic tracking systems~\cite{koivukangas2013technical}, or operate directly on intra-operative images~\cite{Jolesz:Book:2014}. 

Image-based registration in the interventional context has received substantial academic attention~\cite{pluim2003image,liao2013review}. This is because external navigation, while improving surgical accuracy, is associated with increased procedural time,  and complex and manual intra-operative calibration procedures that may lead to a high level of surgeon frustration~\cite{joskowicz2016computer}. It is widely believed that image-based registration will better integrate with procedural workflow, mitigating many negative aspects of external tracking approaches while providing similar accuracy. Further, since no additional hardware is required, there is great potential for widespread adoption and deployment of these purely software-driven methods. This suggests that navigated surgery may also become available in remote and rural hospitals that could not afford dedicated equipment otherwise. 

Despite clear opportunity, image-based registration is not yet widely used in interventional clinical practice. This is because, depending on the clinical context, several challenges of image-based registration have not yet been solved reliably. During surgery, the anatomy undergoes highly complex deformations including the loss of mass or topological changes during resections. Accurately recovering bio-mechanically plausible transformations that represent anatomical change from pre- to intra-operative state that are measured with different imaging modalities is the subject of ongoing research. Here, we will focus on two of the associated challenges: 1) Modeling image similarity between images of the same anatomy but acquired with different modalities, and 2) estimating initial transformation parameters that are good enough for registration algorithms to succeed.

On a high level, image registration seeks to find a transformation that, when applied to the moving image, aligns it with the target image such that locations in both images are in correspondence. Quantifying \emph{correspondence} is achieved using image similarity metrics that, usually, operate on the image intensity values. Straightforward comparison of intensity values, \eg, using 
a simple sum of squared differences,
is generally unrewarding since the underlying assumption on image formation are prohibitively strong, even when moving and target image are acquired with the same imaging modality. For interventional image fusion, the problem is more challenging since images of different modalities must be aligned. In this case, 
the additive Gaussian noise assumption underpinning the sum of squared differences
is certainly violated. Even worse, due to the different physical processes that govern image formation, there is no guarantee that the same anatomical structures are visible in both images, thereby challenging the adequacy of co-occurrence-based similarity metrics, including correlation and mutual information.
Nonetheless, despite these limitations, model-based image similarity criteria currently remain state-of-the-art performers in many interventional image-registration tasks including ultrasound to MRI registration for neurosurgical guidance \cite{Fuerst:MEDIA:2014,Wein:POCUS:2018}.

\subsection{Contextual Learning for Image Registration}
Using deep learning to go past some of the limitation of classical image registration is an active area of research. However, due to the fundamental challenge of gathering ground-truth data for image registration, many of the most successful learning-based registration methods for diagnostic images exploit unsupervised learning and optimise a classical image similarity metric based loss \cite{deVos:MEDIA:2019,Balakrishnan:TMI:2019}.
This approach remain unsuitable for most interventional purposes where more flexible solutions are required.
A prominent example highlighting the need to take the interventional context into account is transrectal ultrasound (TRUS)-guided prostate biopsy. Conventionally, the biopsy target is segmented on pre-operative 3D MR images and this must then be registered to intra-operative 3D TRUS volumes. Since MR and TRUS images exhibit substantially different image appearance, contrast, and artifact level, this suggests that no good mathematical model exists to describe image similarity between these two modalities. Data-driven approaches that do not explicitly model intensity correlations to test for image correspondence but optimize a surrogate measure thereof now achieve state-of-the-art performance. One candidate surrogate measure can be defined by enforcing segmentations of the same structures to exhibit maximal overlap after registration~\cite{Hu:MedIA:2018}.
Remarkably, learning to optimise for such losses does not require access to ground-truth for the spatial transformation and leverages application-specific annotations that are considered as weak annotations.
Further contextual information can be captured by learning data-driven spatial transformation models or regularisation terms
\cite{Hu:MICCAI:2018}.
Related physics-based deformation models have been trained to predict shape changes in segmented organs from sparse annotations which could be used for augmented reality purposes \cite{Pfeiffer:IJCARS:2019,Brunet:MICCAI:2019}.
Taking account of the interventional context one step further, \cite{Hu:MICCAI:2019} noticed that in many cases including MR-TRUS guided biopsy, the main purpose of interventional data fusion is to propagate a patient-specific target defined on a pre-operative image to its interventional counterpart and proposed to replace the registration step by a conditional segmentation one.

Even in scenarios where data-driven similarity metrics may be learned, finding the transformation that optimally aligns a pair of images can remain non-trivial. This is because image similarity is well defined, i.\,e. informative, only in a narrowly circumscribed vicinity around the true transformation, emphasizing the need for appropriate initialization, such that the initial mismatch falls within the \emph{capture range} of the image similarity metric and optimization algorithm~\cite{esteban2019landmarks}. While adequate initialization is challenging in all registration scenarios, it is considered to be most detrimental in slice-to-volume applications. Such applications are common in image-guided interventions, with the most prominent examples being the bijective alignment of 2D B-mode ultrasound to 3D MR or CT volumes or the projective registration of pre-operative 3D MR or CT volumes, or CAD models to intra-operative 2D X-ray or endoscopy images. 

In cases where the 3D imaging protocol context is well defined, i.\,e. one is guaranteed to observe the same extent of anatomy, direct approaches to initialization are possible. These methods only accept the 2D image as input and directly estimate its initial pose relative to a 3D canonical atlas coordinate system that is implicitly defined by the choice of 3D image database~\cite{hou2017predicting,bui2017x} or tool model~\cite{miao2016cnn}. These approaches are attractive, mainly due to two reasons: First, run times are short since only 2D images must be processed; and second, they lend themselves well for scenarios where 2D slices are acquired successively to reconstruct a full 3D volume. However, due to the complexity of the problem and canonical atlas assumption, their performance is often limited in practice.

When a canonical space cannot be defined, alternative approaches typically mimic the external tracking workflow where relative poses are inferred analytically. While external tracking devices require attachment or implantation of artificial fiducial markers to get position information readouts, AI-based approaches seek to establish correspondence directly from the images
or from sparse but corresponding image locations.
In \cite{prevost20183d}, by learning from a dataset of tracked ultrasound, the authors demonstrated that, without inference-time reliance on the tracker, deep learning approaches can estimate the 3D motion occurring in between consecutive 2D ultrasound images with an accuracy far exceeding that of conventional speckle decorrelation techniques and matching that of the external tracker. This is allowing for sensorless 3D freehand ultrasound and creates new opportunities in computer assisted interventions.
Another complementary powerful concept for tracker-less image alignment is the detection and identification of anatomical landmarks. These are particularly appealing since they carry semantic meaning, and consequently, define point correspondence across modalities and domains. Reliably detecting anatomical landmarks is complicated because of changing appearance based on viewpoints, but has recently become possible due to powerful convolutional neural network-based image analysis for anatomical landmarks as shown in the pelvis~\cite{bier2018miccai,esteban2019landmarks} and knees~\cite{bier2018detecting}. The same concept of point correspondence naturally extends to tools and implants where, rather than relying on anatomical landmarks, keypoints on the CAD model are used~\cite{gao2019localizing,kugler2018i3posnet,esfandiari2018deep}. The aforementioned approaches aim at discovering well defined points, however, finding the same arbitrary point in multiple images is equally appropriate to establish correspondence. In this formulation of the problem, an AI-based algorithm is trained to produce a pose invariant latent representation of point appearance. Then, query points can be randomly sampled in one image that are then re-discovered in the target image~\cite{liao2019multiview}, thereby establishing correspondence. This approach is appealing since it does not impose any prior on the imaged object, however, learning a pose invariant latent representation so far has only been demonstrated for comparably small pose differences.


\begin{figure*}[bth!]
\centering
\includegraphics[width=0.7\linewidth]{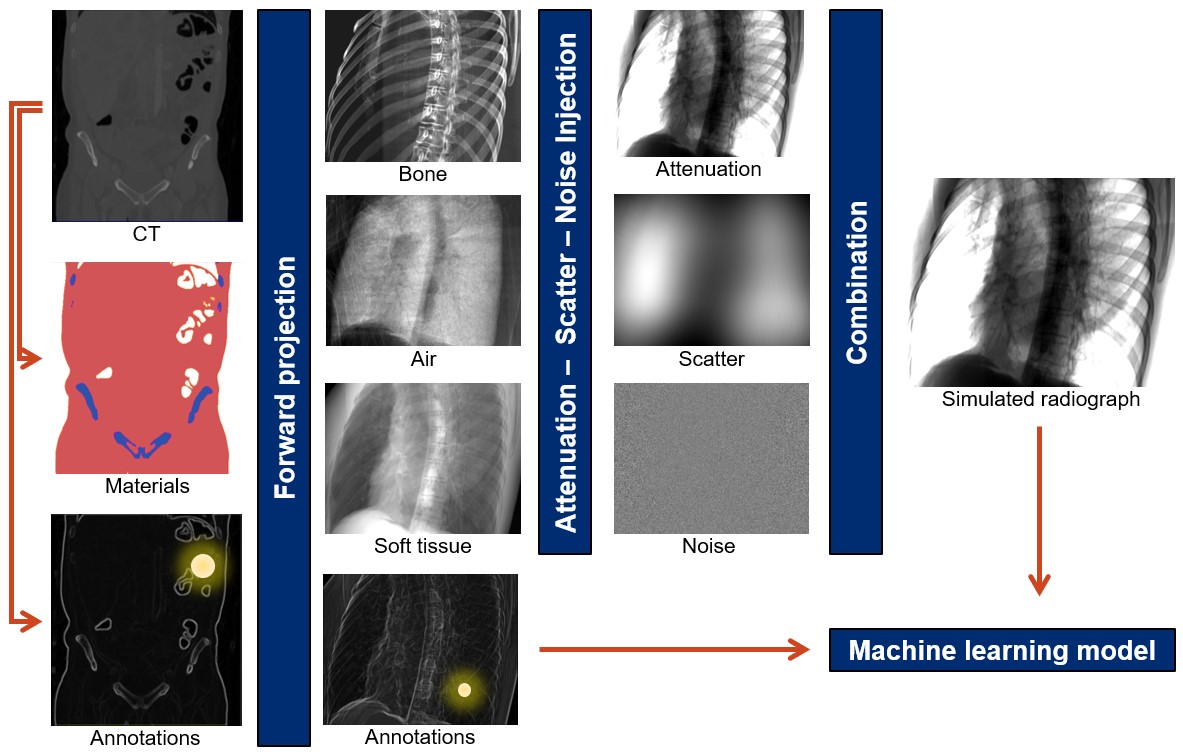}
\caption{Realistic simulation of X-ray image formation from pre-operative CT is one possibility to create large quantities of well annotated images. Pipeline represents the simulation approach described in~\cite{unberath2019enabling}.}
\label{fig:X-ray-simulation}
\end{figure*}

\section{Intelligent Intra-operative Imaging}
\label{sec:modality}
\subsection{From Data Fusion to Intelligent Imaging}
Intelligent intra-operative imaging refers to augmenting the value of intra-operative images for clinical decision making by providing additional information that is tailored to the context of the intervention. In increasingly granular order, context here describes the interventional requirements specific to a certain procedure, step in the surgical workflow, decision, or even surgeon's preferences. So far, efforts in this direction are dominated by data fusion methods that seek to enrich intra-operative images with procedural planning information that exists from pre-operative data. While this approach, even when relying on classical CAI tools, has been deployed successfully for several types of procedures \cite{Jolesz:Book:2014}, it is fundamentally limited in its capabilities of fully leveraging all acquired data. This is because the value of intra-operative images is reduced to a proxy to support, \eg, image-based registration or as a means for overlay, while all \emph{intelligent information} that really augments the decision making is propagated solely from pre-operative images. In addition to under-exploiting intra-operative images, this strategy only allows for displaying information derived from pre-operative data that becomes outdated as surgery progresses. This calls for the development of intelligent intra-operative imaging that fully leverages the information contained in interventionally acquired data in real-time. Augmenting decision making in this way offers clear opportunities by: 1) automating quantitative measurements required for precision medicine; and 2) extracting information that is otherwise not easily accessible which may allow development of new surgical techniques. 
Still, contextual and intelligent interventional image analysis is not yet mainstream technology because, compared to diagnostic image analysis, the environment for developing AI solutions is even more hostile. From our experience working with clinical collaborators across different sites and specialties, we believe that this is primarily due to three reasons.
First, while hundreds of images are acquired for procedural guidance, only very few, if any, are archived \cite{Clunie:JDI:2016,Murad:GIE:2014,Deganello:ULTMED:2018}, thereby suggesting a severe lack of meaningful data for researchers to work with.
Second, learning targets beyond segmentation are not well established or defined.
Third, images of the anatomy are acquired from multiple viewpoints, the exact poses of which are not reproduced nor known.
Finally, the overall variability in the data is further amplified by surgical modification of anatomy and presence of tools. Overall, the accessible data is heavily unstructured and exhibits enormous variation, which challenges meaningful data augmentation strategies. As a consequence, in order to train AI algorithms on interventional images, solutions to the dataset curation and annotation problem must be found first. Overcoming these hurdles seems challenging and is reflected in the observation that only very little work has considered learning in this context. It is worth mentioning that the lack of annotated and/or paired data equally affects other methods presented in this manuscript. 

\subsection{Simulation-based Training}
First steps in addressing the data problem have been taken, serving as a stepping stone for the transformative technology that is \emph{intelligent imaging}. While large scale acquisition of  highly structured data is tractable for some interventional applications, particularly ultrasound~\cite{milletari2019straight,Hu:RAMBO:2017}, most other approaches rely on synthetic data generation from physical models of the scene.
This paradigm is attractive because all quantities of interest are precisely known by design, however, if simulation is performed na{\"i}vely, AI models trained on synthetic data will not generalize to clinically acquired images because of the large domain mismatch paired with poor generalizability of today's models~\cite{unberath2019enabling}.
Three complementary ways have recently been shown to mitigate this problem.
First, if clinically acquired data is available in addition to the well annotated synthetic data, style transfer algorithms can be trained that alter the appearance of real data to close the domain gap, as shown for ophthalmic surgical microscopy \cite{Engelhardt:MICCAI:2018,Luengo:BMVC18}. Using such enhanced simulated data for training of more complex tasks has been applied successfully to endoscopy~\cite{mahmood2018unsupervised} and X-ray imaging~\cite{zhang2018task}.
Second, if too little clinical data is available, learning a style transfer algorithm is impossible. In these cases, a powerful alternative is increasing the realism of synthetically generated images in a model-based approach.
Doing so requires accurate models of all physical principles that govern image formation, however, approximations are usually required to reduce simulation time to acceptable levels. Realistic simulation works well for X-ray-based modalities as illustrated in \figref{fig:X-ray-simulation}) and demonstrated in \cite{unberath2018deepdrr,unberath2019enabling}. It has also been proposed in endoscopic imaging~\cite{mahmood2018deep}. However, the level of required realism likely depends on the application and learning target, since it has been shown that even less realistic simulations could be adequate, \eg, in some ultrasound applications~\cite{nair2018deep}. The aforementioned approaches aim at  reproducing real data appearance which is very complicated in practice. If closely matching real data appearance is found to be impossible, domain randomization can be used to improve the robustness of the trained model to partially unseen data.
Rather than perfectly matching real data characteristics, the goal of domain randomization is to generate multiple versions of the same sample with all but the important characteristics randomized. When training AI algorithms on such datasets, the models are assumed to become robust to these types of domain changes.
Domain randomization can be seen as image formation-based data augmentation and has recently been applied to
X-ray imaging \cite{Toth:MIDL:2019} as well as
colonoscopy~\cite{mahmood2018deep}, where achieving realistic image appearance is very complicated due to fine texture and specular reflectance of the tissue. It is worth mentioning that all the above techniques for synthetic data usage are similar in that AI algorithms never process real data during training. This characteristic is associated with a notable drop in performance when applied to real data due to residual domain mismatch. Consequently, assessing algorithmic performance only on a synthetic test set will severely overestimate the AI models accuracy during deployment and quantitative experiments on clinical data are required. Ultimately, training the AI directly on real data is preferable, highlighting the need for further research on un- and self-supervised learning to leverage large quantities of  unlabeled data.

\subsection{Intelligent Imaging in Interventional Biophotonics}
Although conventional interventional imaging, such as X-ray fluoroscopy, surgical microscopy, endoscopy and ultrasound will benefit from being augmented by contextual AI, another interesting area in which the intelligent imaging paradigm is expected to make an important impact is that of interventional biophotonics imaging. The initial focus in biophotonics has been on developing optimal, task-specific, contrast agents that would be merely be directly visualised, e.g. in tumour-specific fluorescence imaging. The biophotonics community has however faced stringent challenges in identifying versatile contrast agents suitable for use in patients and realised that tissue differentiation would remains challenging with such an approach. Advanced high-dimensional optical imaging techniques are currently seen as promising solutions for intraoperative tissue characterisation, with the advantages of being non-contact, non-ionising and non- or minimally-invasive. However, because of the high-dimensional nature of the generated data, direct visualisation by the clinical team becomes impractical. This calls for automated learning-based information extraction before display.
As in the previous examples of intelligent imaging, many of the most advanced AI-supported interventional biophotonics imaging devices currently exploit model-based learning or unsupervised learning.
Point-based measurement devices able to measure Raman scattering have recently been translated into commercial products \cite{Jermyn:SCITMED:2015} with support from supervised classification \cite{Jermyn:JBO:2016} or usupervised dimensionality reduction \cite{Banbury:SCIREP:2019}.
Addressing the lack of wide-field information in point-based systems, the community has looked into modalities such as hyperspectral imaging \cite{Shapey:JBIO:2019}
with an increasing use of machine learning to solve some of the intrinsic challenges of high-dimensional data. Indeed, while bearing rich information, the raw 2D-space+wavelength+time data that hyperspectral imaging produces is difficult to interpret for clinicians as it generates a temporal flow of three-dimensional information which cannot be simply displayed in an intuitive fashion.
Innovative use of Invertible Neural Networks in combination with model-driven simulation has been used to train neural network based regressors which are capable of real-time operation and can provide uncertainty estimates for oxygen saturation measurement from hyperspectral data \cite{Adler:IJCARS:2019}.
Unsupervised deep manifold embedding for hyperspectral imaging was proposed in \cite{Ravi:TMI:2017} and deep learning was used for reconstruction from sparse hyperspectral data~\cite{Lin:MEDIA:2018}.
Intelligent imaging concept with simulation- or model-based training are also being progressed with other emerging biophotonics imaging modalities such as for super-resolution in endomicroscopy~\cite{Ravi:MedIA:2019,Szczotka:IJCARS:2018},
and artefact suppression in photoacoustic imaging \cite{Allman:IUS:2018}.

\subsection{Towards Prospectively Planned Intelligent Imaging}
With the availability of training data,
either via dedicated data collection or synthetic generation, AI algorithms can be developed to analyze intra-operative images in near real-time and supply contextual information to improve decision making.
Omitting applications to endoscopic video sources which are discussed in depth in \secref{sec:surgvision}, and focusing first on interventional X-ray imaging,
benefits of real-time machine learning range from segmentation of tools~\cite{ambrosini2017fully,breininger2018multiple,gao2019localizing}, anatomical landmark detection~\cite{bier2018miccai,bier2018detecting}, anatomy localization~\cite{sa2017intervertebral},
and denoising~\cite{matviychuk2016learning,Hariharan:MICCAI:2019},
to surgical phase recognition~\cite{ambrosini2017fully}.
Corresponding developments can be found for ultrasound imaging~\cite{anas2018deep,Mwikirize:IJCARS:2018,Pesteie:TMI:2017}.

While the above lists of applications merely hints at the potential that AI-based analysis of internventional images has to offer, there is an interesting observation: The majority of \emph{intelligent imaging} algorithms, including all aforementioned methods, try to provide richer information by automated analysis of traditionally acquired images, with little or no knowledge of the image acquisition workflow. This raises an interesting question: If it is known what information is desired or desirable at any given point during surgery, is it possible to prospectively acquire an image that is most informative in that particular context? First steps in this direction have recently been reported, exploiting ultrasound image formation to suppress scatter~\cite{luchies2018deep} or beamforming a B-mode image~\cite{simson2019end,simson2018deep} together with producing its segmentation~\cite{nair2018deep}. Zaech et al.~\cite{zaech2019taskaware} use an AI-based algorithm to recommend task-optimal and patient specific C-arm X-ray trajectories during cone-beam CT of spinal fusion surgery, and similar ideas arise for ultrasound transducer positioning~\cite{milletari2018cfcm}. 

The domain of real-time interventional image analysis is fairly untapped as of yet but offers great opportunities for workflow analysis, surgical progress monitoring including anticipation and adverse event detection, and supplying rich information for human-in-the-loop decision making. Additionally, task-aware and autonomous imaging modalities may benefit interventional imaging already one step before the image is analyzed and may thus give rise to disruptive technology and novel surgical approaches.

\section{Surgical and Endoscopic Vision}
\label{sec:surgvision}

\subsection{Recognizing Endoscopic Activity}
Standard endoscopic imaging is certainly the modality most closely relating to natural images. It should therefore not be surprising that machine learning tools for interventional images have developed most rapidly in this field.
As a proxy for the eyes of the surgeon inside the patient, the endoscopic camera is the privileged source of digital information to understand the activities performed during endoscopic procedures. Endoscopic videos usually capture most of the activities performed within the patient.
Recognizing and understanding these activities is essential to develop novel assistance systems that are reactive to the context, \eg, that can provide timely instructions to OR staff, enforce safety checkpoints or log automatically relevant information within the surgical report. Surgical activity recognition from endoscopic videos is however a highly challenging task due to the variability existing across patients, surgical treatments and surgical teams. 

In the recent years, a large body of work has focused on recognizing the surgical steps of a procedure directly from the videos \cite{blum:miccai2008,Lalys2012,Quellec2014a,Twinanda2017,Volkov:icra17,Bodenstedt:ijcars19}. This has notably been the case in cholecystectomy, a common procedure consisting in removing the gallbladder, which is frequently used in research due to its high frequency of occurrence and well-standardized protocol~\cite{Padoy2012}. There, the steps include for instance ``calot triangle dissection, cystic duct and artery clipping and cutting, gallbladder dissection and gallbladder packaging". Recognition of these steps allows for the automated understanding of the progress of the surgery.
To perform recognition, models of the underlying workflow of the procedure are learnt from datasets of exemplary videos, annotated manually with the different steps. In \cite{Twinanda2017}, the model consists for example of a visual feature extractor relying on a deep neural network that feeds a temporal recognition model, like a hierarchical hidden Markov model or an LSTM model. Several types of procedures have been successfully studied for step recognition besides cholecystectomy. Examples are cataract surgery \cite{Lalys2012,Quellec2014a} and laparoscopic sleeve gastrectomy \cite{Volkov:icra17}. As the current recognition methods show very promising results and real-time capabilities, they can potentially be directly embedded in the endoscopic tower to deliver contextual support. Other interesting prediction tasks have been tackled with success using deep learning methods. In \cite{Twinanda-rsd18,bodenstedt:ijcars2019rsd}, the remaining  duration of the procedure is predicted in real-time using deep recurrent models trained directly from video data. In \cite{Twinanda2017, Hajj-media18, Nwoye2019}, the presence of the instruments in the surgical scene is automatically detected. Additional applications include bleeding and smoke detection \cite{Leibetseder:MMM18, okamoto:ijcars2019}, as well as surgery type identification at the beginning of the procedure \cite{Kannan:tmi2019}.

Beyond the recognition of the surgical steps indicating the progress of the surgery and the recognition of events such as bleeding, many potential applications, like safety monitoring and human-robot cooperation, require a finer level of understanding of the surgical activities. 
Future research therefore needs to demonstrate accurate recognition of the detailed interactions between the tools and the anatomy. To have impact beyond a single operating room, recognition methods will also need to scale up to different types of surgeries, operating rooms and hospitals without requiring the manual annotations of large datasets for each situation. Recent methods exploiting non-annotated videos through self-supervision or weak-supervision \cite{Funke:or20:2018,yengera:arxiv2018,ross:ijcars2018, Yu:ipcai2019, Nwoye2019} or exploiting synthetically generated surgeries \cite{Luengo:BMVC18} may prove very useful to train the next generation of surgical recognition systems.

\begin{figure*}[tb]
\centering
\includegraphics[width=0.95\textwidth]{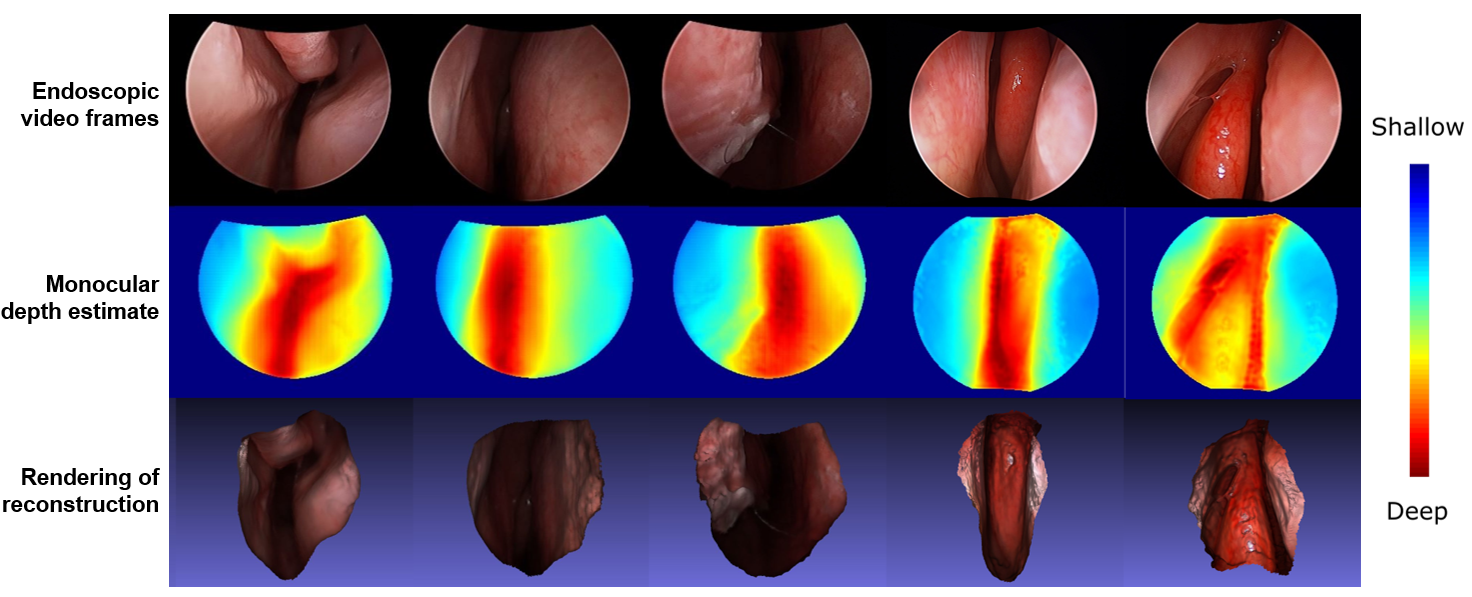}
\caption{Endoscopic video (top), monocular depth estimate (middle), and rendering of a photorealistic reconstruction (bottom). Results were achieved using the self-supervised method described in \cite{Liu:OR2.0:2018}.}
\label{fig:monocular_depth}
\end{figure*}

\subsection{Understanding Image Semantics}
Understanding the surgical scene from the endoscopic images is fundamental for context-aware intelligent computer-aided assistance. During augmented reality visualization, precise pixel-based segmentation of the tools is necessary for handling occlusions and providing the user with the correct perception. Implementing safety warnings, such as no-go zones, requires the detection of the critical anatomy. When another imaging modality is used, its registration to the endoscopic video may require the localization of anatomical landmarks~\cite{Ntourakis:wjs2015}. Similarly, implementing degrees of autonomy during robotic surgery requires the localisation and recognition of the neighboring tools and anatomy. 

Recently, a large body of work has targeted the detection and segmentation of surgical instruments \cite{bouget2017vision}. Deep learning methods have been proposed for both  bounding box or articulated tool detection \cite{sarikaya2017detection,Kurmann:miccai2017,Yin:wacv2018} and for pixel-based tool segmentation \cite{Peraza-Herrera:IROS:2017,laina2017concurrent}. Their superiority has been confirmed on laparoscopic and surgical microscopy datasets in two international challenges organized in 2015 and 2017 at the MICCAI conferences \cite{Bodenstedt:arXiv:2018,Allan:arxiv2019}. Still, the datasets used for evaluation are limited in size and variability. They are far from representing the diversity of surgical scenes, which can indeed be very challenging due to the presence of occlusions, smoke, bleeding, specularity, motion blur, and deformation. Furthermore the aforementioned approaches are fully-supervised and therefore impose important burden on the collection of representative training datasets. New approaches are needed that can generalize easily to various types of procedures and be trained using weaker information for training, such as image-level tool presence \cite{Nwoye2019}, point annotation \cite{Lejeune:media18} or scribbles \cite{Fuentes-Hurtado:ijcars19}.

Far less work has addressed the much needed anatomy detection and segmentation, certainly due to the lack of available public datasets. The community is however putting large efforts in this direction, as illustrated by the recent generation of the CaDIS dataset \cite{Flouty:arxiv2019}, which contains pixel-level annotations for 36 semantic classes in cataract surgery videos. Progress has also been achieved in specific areas, such as
liver segmentation \cite{Gibson:SPIEMI:2017},
lesion detection and characterisation during gastroscopy  \cite{Everson:UEGJ:2019}  or polyp detection during colonoscopy \cite{Misawa:gastro2018,Wang:GUT:2019}. Here again, deep learning is the state-of-the art, as demonstrated for polyp detection in a challenge organized at MICCAI 2015 \cite{Bernal:TMI:2017}. Thanks to the real-time capabilities of deep learning approaches, the intra-operative benefits of such systems already start to be evaluated in randomised clinical trials~\cite{Wang:GUT:2019}.

\subsection{Reconstructing Anatomic Geometry}
 Endoscopy mimics the surgeon's eyes within the body, but due to the monocular construction of endoscopes it lacks one important visual cue: Depth. This shortcoming has implications: It has recently been shown that the availability of 3D anatomic geometry benefits several clinical tasks, including the detection of critical anatomy such as polyps~\cite{mahmood2019polyp} and the registration of pre-operative 3D data to endoscopy video to enable navigation~\cite{leonard2018evaluation}. In addition, analyzing 3D representations of anatomy would allow for the introduction of quantitative measurements, enabling the standardization of clinical reporting across sites. Recovering anatomic 3D geometry, \eg to augment endoscopic video with depth cues or to provide dense 3D reconstruction, has gained considerable traction and is now an emerging discipline with developments often orthogonal to those for complementary tasks \eg segmentation. This is because deep learning-based algorithms are able to exploit image-level features to provide dense depth estimates even from monocular video, complementing traditional optical endoscopy with depth sensing as ''pseudo modality''. However,  training depth estimation algorithms on endoscopic sequences is complicated in practice because no paired depth measurements exist naturally. While paired data can be generated in silico via simulation from CT~\cite{mahmood2018unsupervised,mahmood2018deep,visentini2017deep}, the resulting trained models will need to overcome the domain mismatch to real clinical data with methods similar to that presented in \secref{sec:modality}.
Recently, self-supervised training paradigms that rely on traditional multi-view stereo approaches have received increasing attention as they can be trained directly and solely from endoscopic video. Multi-view stereo algorithms including structure from motion~\cite{leonard2018evaluation,Liu:OR2.0:2018} and simultaneous localization and mapping~\cite{wang2018surgical} can be adapted to work with endoscopic video, but  they cannot provide dense 3D reconstructions due to the lack of photometric constancy in endoscopic video and texture scarceness that complicate feature matching across frames. These algorithms do, however, provide a few reconstructed 3D points, and more importantly, relative camera poses that can be used to supervise monocular depth estimation~\cite{wang2018surgical,Liu:OR2.0:2018}.
A representative photorealistic reconstruction achieved using a structure from motion supervised depth estimation method is shown in \figref{fig:monocular_depth}.
These methods achieve state-of-the-art performance with good generalization ability, however, the resulting reconstructions are only up to scale. 
Among the biggest premises of video-based reconstruction is the possibility of monitoring anatomical change during surgery. This would require methods to robustly handle various sorts of uncontrollable variation, including bleeding, smoke, or tool presence. Solutions to these problems are currently unknown. Even in more controlled scenarios, widespread adoption of learning-based reconstruction from endoscopic video is hindered by the lack of publicly available datasets, making it unclear how well today's algorithms perform on clinical data. This challenge is further aggravated by the lack of direct evaluation targets: When applied to real clinical data, current reconstruction or dense estimation algorithms can only be evaluated via surrogate tasks, such as video-CT registration~\cite{Liu:OR2.0:2018,liu2019self} or polyp classification~\cite{mahmood2019polyp}.

\section{Clinical Workflow Monitoring and Support}
\label{sec:datasci}

\begin{figure*}[tb]
\centering
\includegraphics[width=0.8\textwidth]{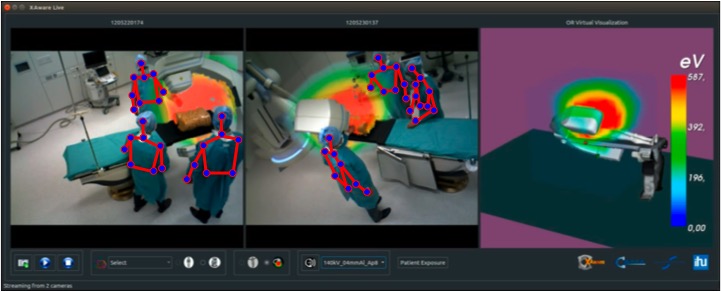}
\caption{Capturing the 3D context of the operating room is necessary for providing AI-based decision support and monitoring risk. In this example, the staff radiation exposure during a X-ray based procedure is computed in-situ via simulation and displayed with augmented reality in a training scenario \cite{Rodas2017}.}
\label{fig:AI-radiation}
\end{figure*}

\subsection{The Notion of Surgical Control Tower}
While imaging alone provides valuable information,
modern procedures rely increasingly on a variety of complex devices and intricate workflows.
This limits the knowledge extraction that AI systems can do based on imaging alone,
and makes it difficult for humans to properly analyse in real-time the wealth of available data.
Furthermore, even though the quality of care has generally improved with the introduction of new surgical techniques and devices, adverse events still occur, a large part of which are preventable \cite{James2013,Suliburk:jama2019}. Humans are prone to fatigue, teams to miscommunications, devices can fail, and for all roles, surgical tasks require an ever increasing level of specialization. 
The increased use of digital equipment in the OR however opens up new opportunities for support and monitoring, at the level of the whole room, by providing artificial intelligence systems with real-time data that capture a faithful representation of the processes taking place during the surgery. Indeed, most of the activities happening in the room can be captured digitally either through interactions with equipment, such as information systems, room control interfaces, imaging devices and instruments, or though the use of sensors, such as ceiling-mounted cameras, which are now becoming widespread and increasingly used for documentation, teaching and augmented reality assistance. 
Consequently, it is highly likely that in the near future assistance systems will be fully integrated in a digital OR that will monitor surgical processes though AI, akin to a \emph{surgical control tower}~\cite{Padoy2018,Padoy2019}, that can analyze the whole digital information in real-time to provide context-aware support and information within and outside the OR.
Applications for such a control tower are for instance the transmission of live information about the OR status, the adaptation of user-interfaces to the surrounding context, the display of instructions within the OR, the creation of an automated report, the recording of the activities for archiving and legal purposes, the enforcement of safety checklists, the detection of anomalies with respect to past workflows, and improved scheduling for staff and patients. To perform these tasks, the control tower will have access to and crunch masses of multi-modal digital data coming from hundreds of past surgeries.

\subsection{An Endeavour Rooted in Surgical Data Science}
An essential component of the control tower is the data-driven modeling and understanding of the clinical activities, an undertaking that taps into the emerging research field of Surgical Data Science~\cite{Maier-Hein:NATBME:2017,Vedula:ISS:2017}.
Machine learning has been key to generate models of procedural interventions from data \cite{Jannin:neuro2007,Blum:ijcars2018} and ontologies have also been developed to standardize the resulting models \cite{Gibaud:ijcars2018}. 
Implementations of such AI-based applications start to emerge in various institutions, besides the ones focusing on analysing endoscopic videos already mentioned in \secref{sec:surgvision}. As video data remains one of the main source of information, they highly rely on deep learning. Videos captured by cameras mounted in the room provide indeed a rich source of information about the activities without disrupting the workflow. 
For instance, a patient and staff radiation exposure monitoring system for hybrid procedures illustrated in \figref{fig:AI-radiation} was proposed in \cite{Rodas2017}. It relies on several RGB-D cameras to estimate the 3D pose of the persons and room layout, which can then be used to simulate and visualize in situ X-ray propagation around the patient table. \cite{Haque2017mlhc} develops a system to monitor hand-hygiene in hospital corridors in order to analyse and reduce hospital acquired infection. The approach uses a large set of depth cameras installed to observe the hand-soap dispensers. For the intensive care unit, \cite{Ma:ccm2017,Yeung2019npj} present methods based on color or depth video data for the detection of patient mobilization activities. 
Key building blocks to the success of these applications are the estimation of clinician and staff poses \cite{Haque2016,Belagiannis:mva2016, Kadkhodamohammadi2017}, as well as the recognition of their activities \cite{Chakraborty2013,Lea2013,Twinanda2015,Twinanda2016}. As for traditional visual data, deep learning based approaches are currently the best performing methods for these tasks, though it should be noted that they do not necessarily perform as well on clinical data yet. This is due to the specificity of clinical videos, where staff wear gowns and masks, colors are often similar, and cameras observe the room from restricted positions, but also from the fact that there is no clinical COCO or Imagenet dataset yet. \cite{srivastav2018mvor} evaluates state-of-the-art human pose estimation approaches and \cite{Issenhuth2019} state-of-the-art face detection approaches on clinical data. Both studies show a large margin for improvement. Since the development of large annotated datasets of clinical videos may be difficult due to the expertise required and the restrictions on data, other approaches need to be developed, for instance using non annotated data for transfer learning \cite{Issenhuth2019}. 

This will also help deploy the surgical control tower in new clinical environments, as the variability in room layout, camera configuration and workflow can be high. Retraining the assistance systems using only non-annotated data from the novel environment or a tiny subset of annotated data will be crucial for the adoption of these technologies. As even the collection of non-annotated video data can be challenging due to data and privacy regulations, it may also be required to implemented federated learning approaches or develop methods able to cope with privacy-preserving data, such as depth-only videos \cite{Haque2017mlhc} or even low-resolution depth videos \cite{srivastav:miccai2019}. In \cite{srivastav:miccai2019}, it is shown that 2D human pose estimation can be achieved with reasonably high accuracy on depth images downsampled by ten to the resolution 64x48. By using other information, such as system events \cite{Malpani2016} or speech analysis \cite{Gu2017}, the analysis of clinical activities will be further improved.

\section{Discussion and Conclusion}%
\label{sec:outlook}
While AI is starting to impact CAI, as described in this paper,
there are a number of challenges that are specific to surgery and intervention to overcome to deliver clinical impact. Leveraging context within learning paradigms will be crucial to address those in a clinically meaningful way.
The emerging field of CAI4CAI offers researchers a large set of open problems to tackle.
These notably stem from
the heterogeneity of surgical procedures and their particular requirements for intra-operative imaging~\cite{navab2016personalized};
the difficulties in data acquisition;
the complexity in modeling and inferring decision making processes;
and the intricacy of the execution of surgical tasks.
Over the years, the CAI community has defined increasingly powerful Surgical Process Models \cite{Lalys:IJCARS:2014} to
gain actionable understanding of surgical procedures while describing interventions as a sequence of tasks and activities at different granularity levels.
At the finest level, mapping what should be the \emph{Language of Surgery} \cite{Reiley:IEEEspectrum2011}, researchers currently break down surgical gestures into semantically relevant motion units called \emph{surgemes} that are further composed of sequences of motion primitives named \emph{dexemes} \cite{lin:miccai2005,Lin:CAS:2006,Despinoy:TBME:2015}.
Yet, this taxonomy mostly focused on the surgical action and in particular on surgical tool manipulation
and could thus rather be considered as mapping the \emph{Language of Surgical Dexterity}.
This is already a laudable achievement
and led to scientists and engineers being able to, \eg, quantify the success of a training program for executing different surgical actions \cite{Varadarajan:miccai2009,padoy:icra2011}.
As suggested by the study conducted by Birkmeyer et al. for bariatric surgery \cite{Birkmeyer:nejm2013}, surgical skills can be highly correlated to surgical outcome for certain procedures. AI systems have been shown capable of evaluating technical skills using data from either training scenarios \cite{Zia:ijcars18} or real procedures \cite{Kim:ijcars2019}.
Yet, by severely under-utilising the rich information contained in other data sources,
the \emph{Language of Surgical Dexterity}
is still not capturing the most complex aspects of surgical decision making.
To address the need to capture, understand and support all the cognitive interactions and processes taking place in the operating room, the Surgical Data Science community will need to drive the deployment of real-time multi-modal data acquisition systems that will be used routinely. At the same time, it will foster the development of new standards and regulations aiming at increasing the interoperability of data, devices and models. This will directly benefit CAI4CAI by simplifying the implementation and training of learning algorithms involving databases from multiple institutions while maintaining privacy, e.g. through federated learning.
CAI4CAI in combination with Surgical Data Science and Surgical Process Modelling could thus aim at defining and understanding the ultimate \emph{Language of Surgery} based on a large number of heterogeneous data sources used continuously by surgeons and interventional teams to guarantee the best outcomes for a given procedure.
As the field blossoms, CAI4CAI researchers will address some of the most rewarding questions in computer assisted intervention.
Could CAI4CAI allow us to learn how decisions are made, or missed, throughout surgical procedures? Could CAI4CAI support such decision makings? Instead of going through the traditional path of segmentation, registration, navigation and visualization, could contextual machine learning allow us to optimize these steps for each given objective and allow for real-time computation and feedback based on large amount of heterogeneous data including pre- and intra-operative imaging, patient characteristics and surgeon preferences?

With more capable and flexible learning paradigms, synergistic collaboration is expected to happen between humans and AI-powered actors.
The field is already seeing exciting attempts to bring the user and the user experience at the centre of our research questions.
For example, novel spatially-aware visualisation beyond traditional user interfaces is explored in \cite{Rodas2017,FotouhiUSGJOAN19}.
The challenge of improving human situational awareness in operating rooms with solutions beyond visualisation is addressed in \cite{Matinfar:IJCARS:2018} with the use of context-specific soundtracks.
Introduction of novel multimodal interaction paradigms and technologies within operating rooms will require extensive use of machine learning to optimize the user interfaces and to provide maximally relevant information and support, while preventing inattentional blindness~\cite{pauly15}.
By developing  systems able to learn from previous surgeries performed by experts how to provide context-aware support and instructions directly in the OR, in the manner of a virtual coach as in \cite{Malpani:arxiv2017}, AI could have a strong impact in improving patient care.
This is another aspect of CAI4CAI which needs particular focus from the scientific community and requires multidisciplinary teams including clinicians, user experience experts and machine learning scientists to work together and come up with intelligent end-to-end CAI solutions.

Finally, in this paper we did not have a particular focus on robotics. However both surgical robotics and robotic imaging will play increasingly crucial roles in the years to come.
Machine learning is demonstrating convincing results in real-time tool tracking~\cite{reiter2012feature,du:tmi2018,pakhomov2017deep,Peraza-Herrera:IROS:2017}. This for example enables automatic positioning of intra-operative OCT imaging planes within surgical microscopy for ophthalmic surgery \cite{laina2017concurrent, rieke2016real}. Integration of robotics within surgical suites would require them to act intelligently, synergistically with the human team and to be fully context-aware at all moments. The wish to have  real-time multi-modal imaging requires full intelligence and  automation. It also requires direct communication and collaboration between surgical robots, imaging robots, surgeons and surgical teams. CAI4CAI will have the challenge of enabling such ultimate intelligence, which requires many years of research and development in many disciplines while remembering past experience
with the first generation of context-aware computing \cite{Erickson:CACM:2002}.
Not only does CAI4CAI offer numerous exciting research directions but it also promises to revolutionize surgery and therefore the future of healthcare at a global scale.

\section*{Acknowledgment}
Tom Vercauteren is supported by a Medtronic / Royal Academy of Engineering Research Chair [RCSRF1819\textbackslash7\textbackslash34], by the Wellcome Trust [203148/Z/16/Z; WT101957] and by the Engineering and Physical Sciences Research Council (EPSRC) [NS/A000049/1; NS/A000027/1].
Mathias Unberath is supported by an NIH/NIBIB R01 EB0223939, Johns Hopkins University internal funding sources, and a Fellowship of the Malone Center for Engineering in Healthcare at Johns Hopkins University. 
Nicolas Padoy is supported by French state funds managed by the ANR within the Investissements d'Avenir program under references ANR-16-CE33-0009 (DeepSurg), ANR-18-CE45-0011-03 (OptimiX) and ANR-11-LABX-0004 (Labex CAMI) and by BPI France (project CONDOR).

\bibliographystyle{IEEEtran}
\bibliography{%
   jn_abrv,%
   bib_data_fusion,%
   bib_intelligent_intra_op_imaging,%
   bib_intervention_planning,%
   bib_misc,%
   bib_surgical_vision_endoscopy,%
   bib_tool_tracking,%
   bib_workflow,%
   ieeebib_settings
   }

\begin{IEEEbiography}%
[{\includegraphicsfilled{1in}{1.25in}{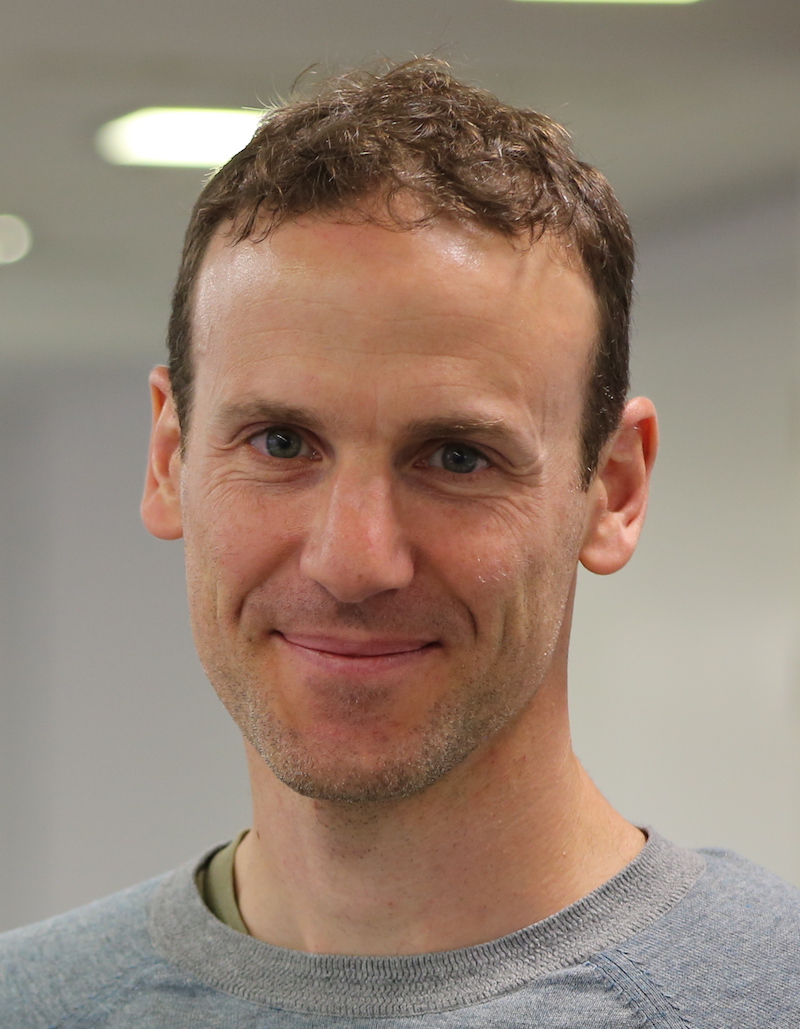}}]{Tom Vercauteren}
Tom Vercauteren is Professor of Interventional Image Computing at King’s College London since 2018 where he holds the Medtronic / Royal Academy of Engineering Research Chair in Machine Learning for Computer-assisted Neurosurgery. From 2014 to 2018, he was Associate Professor at UCL where he acted as Deputy Director for the Wellcome / EPSRC Centre for Interventional and Surgical Sciences (2017-18). From 2004 to 2014, he worked for Mauna Kea Technologies, Paris where he led the research and development team designing image computing solutions for the company’s CE-marked and FDA-cleared optical biopsy device.
He is a Columbia University and Ecole Polytechnique graduate and obtained his PhD from Inria in 2008.
Tom Vercauteren's research focuses on translational medical image computing, machine learning and interventional imaging devices with a specific interest in their development for surgery and interventional sciences.
\end{IEEEbiography}

\begin{IEEEbiography}%
[{\includegraphicsfilled{1in}{1.25in}{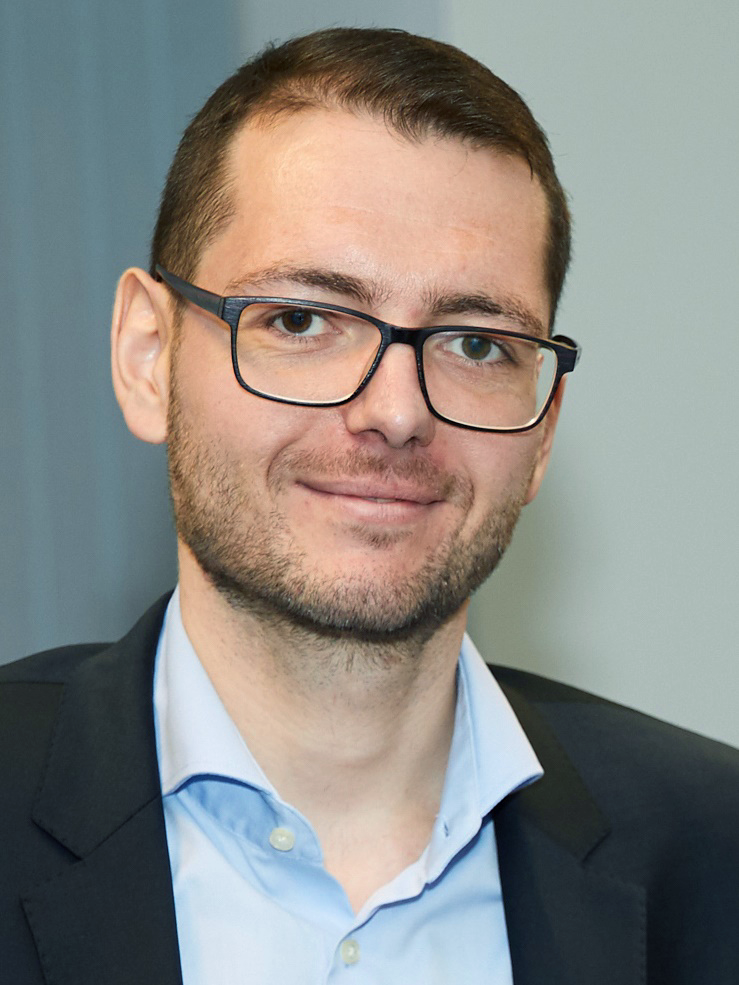}}]{Mathias Unberath} is an Assistant Research Professor in the Department of Computer Science at Johns Hopkins University, and is affiliated with the Laboratory for Computational Sensing and Robotics and the Malone Center for Engineering in Healthcare. Mathias first joined Hopkins as a postdoctoral fellow after graduating summa cum laude from the Friedrich-Alexander-Universit{\"a}t Erlangen-N{\"u}rnberg with a BSc in Physics, a MSc in Optical Technologies, and a PhD in Computer Science. He was an ERASMUS scholar at the University of Eastern Finland and DAAD fellow at Stanford University. Mathias’ research at the intersection of computer vision including augmented reality, machine learning, and medical physics has been recognized with multiple national and international awards, and aims at pushing the boundaries of computer assistance in medical imaging and image-guided interventions. 
\end{IEEEbiography}

\begin{IEEEbiography}%
[{\includegraphicsfilled{1in}{1.25in}{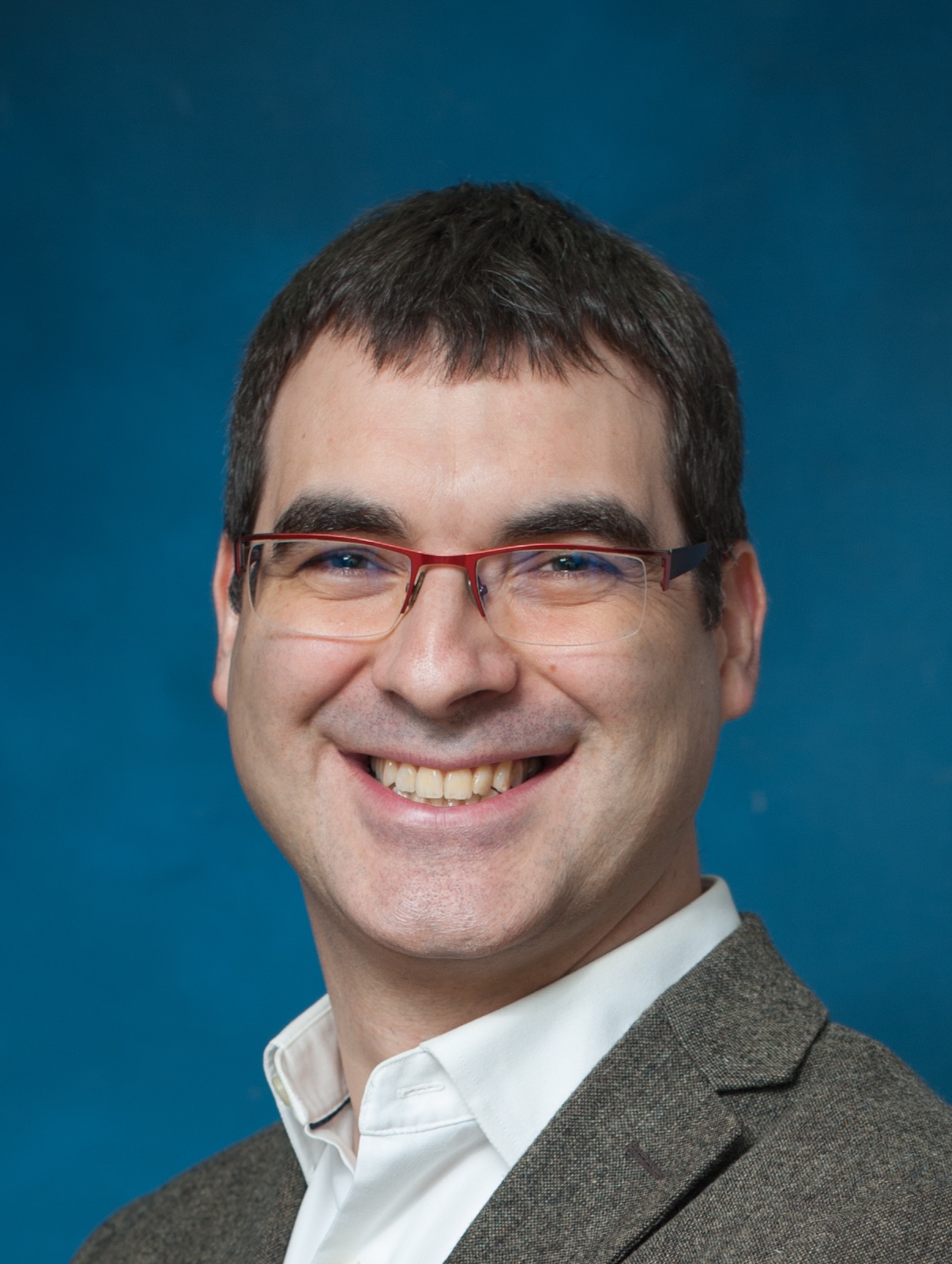}}]{Nicolas Padoy}
is a full Professor of Computer Science at the University of Strasbourg, where he began as an Assistant Professor on a Chair of Excellence in 2012. He created and is currently leading the research group CAMMA on Computational Analysis and Modeling of Medical Activities, which focuses on computer vision, activity recognition, artificial intelligence and the applications thereof to surgical workflow analysis and human-machine cooperation during surgery. From 2009 to 2011, he was a postdoctoral researcher and later an Assistant Research Professor in the Laboratory for Computational Interactions and Robotics at the Johns Hopkins University. He completed his PhD jointly between the Chair for Computer Aided Medical Procedures at Technische Universit\"at M\"unchen (TUM) and the INRIA group MAGRIT in Nancy. He graduated with a Ma\^itrise in Computer Science from the Ecole Normale Sup\'erieure de Lyon in 2003 and with a Diploma in Computer Science from TUM in 2005. 
\end{IEEEbiography}

\begin{IEEEbiography}%
[{\includegraphicsfilled{1in}{1.25in}{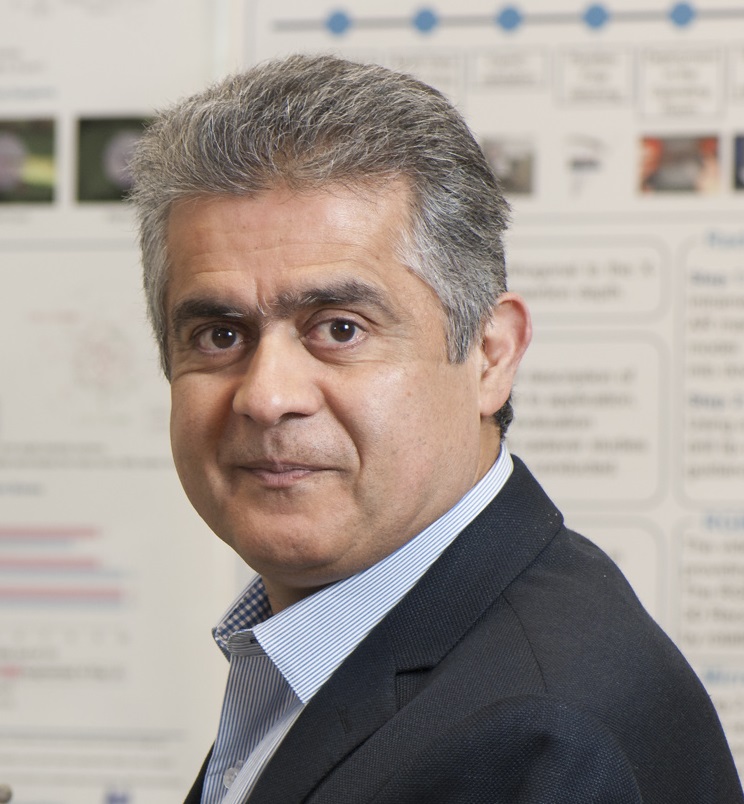}}]{Nassir Navab}
is a full Professor and Director of the Laboratory for Computer Aided Medical Procedures, Technical University of Munich and Johns Hopkins University.
He completed his PhD at INRIA and University of Paris XI, France, and enjoyed two years of a post-doctoral fellowship at MIT Media Laboratory before joining Siemens Corporate Research (1994-2003). He  received the Siemens Inventor of the Year Award in 2001, the SMIT Society Technology award in 2010 and the ‘10 years lasting impact award’ of IEEE ISMAR in 2015.
In 2012, he was elected as a Fellow of the MICCAI Society. He has acted as a member of the board of directors of the MICCAI Society, 2007-2012 and 2014-2017, and serves on the Steering committee of the IEEE Symposium on Mixed and Augmented Reality (ISMAR) and Information Processing in Computer Assisted Interventions (IPCAI).
He is the inventor of 47 granted US patents and more than 50 International ones. His current research interests include multimodal imaging, medical augmented reality, computer assisted surgery, medical robotics, and machine learning.
\end{IEEEbiography}

\end{document}